\documentclass[useAMS,usenatbib]{mn2e}
\usepackage[dvips]{graphicx}
\usepackage{epsfig}
\newcommand{\mnras}{MNRAS}

\newcommand{\D}{\displaystyle}

\title[Short-term Evolution of Initially Nearby Orbits]{Miller's
instability, microchaos and the short-term evolution of initially
nearby orbits}

\author[A. Helmi \& F. G\'omez]{Amina Helmi\thanks{Email:ahelmi@astro.rug.nl} 
\& Facundo G\'omez \\
Kapteyn
Astronomical Institute, University of Groningen, P.O. Box 800, 9700 AV
Groningen, The Netherlands}

\begin{document}

\date{}

\pagerange{\pageref{firstpage}--\pageref{lastpage}} \pubyear{}

\maketitle

\label{firstpage}

\begin{abstract}

We study the phase-space behaviour of nearby trajectories in
integrable potentials. We show that the separation of nearby orbits
initially diverges very fast, mimicking a nearly exponential
behaviour, while at late times it grows linearly. This initial
exponential phase, known as Miller's instability, is commonly found in
N-body simulations, and has been attributed to short-term
(microscopic) N-body chaos.  However we show here analytically that
the initial divergence is simply due to the shape of an orbit in
phase-space. This result confirms previous suspicions that this
transient phenomenon is not related to an instability in the sense of
non-integrable behaviour in the dynamics of N-body systems.

\end{abstract}

\begin{keywords}
stellar dynamics -- methods: analytical -- methods: N-body simulations
-- galaxies: kinematics and dynamics
\end{keywords}

\section{Introduction}

The problem of how exactly galaxies reach their final equilibrium
configuration is still unsolved.  It is clear that, unlike for gases,
two-body collisions between stars in galaxies are not the driving
mechanism to reach a relaxed state, since the associated timescales
are exceedingly large \citep{bt}. In an attempt to explain the road to
equilibrium from a statistical mechanics point of view, Lynden-Bell
(1967) introduced the concept of ``violent relaxation''. In this
context, the relaxation is reached through the effects of a
``violently changing'' gravitational field. However, the detailed
physics of this process also remain to be understood
\citep{arad,valluri}.

Besides the statistical mechanics approach, it is also possible to
study the problem of ``relaxation'' at the level of orbits. In this
case, it is useful to introduce the concept of mixing, by which we
mean how quickly nearby particle trajectories diverge in (phase-)space
as a function of time. In the case of time-independent gravitational
potentials it is customary to classify mixing into two types. If the
particles move in an integrable potential, nearby orbits will diverge
as a power-law in time, e.g. \citet{hw}. This process is known as
phase-mixing \citep{bt}. However, when the potential admits a certain
amount of chaos, there exist regions of phase-space where nearby
orbits diverge exponentially, evidencing an extreme sensitivity to
small changes in the initial conditions \citep{Arnold}. This process
is known as chaotic-mixing \citep{Kandrup-Mahon,Kandrup}.

These mixing processes can also take place in a time-dependent
gravitational potential, in which case the energies of the particles
will not be constant. The degree of ``stickiness'', quantified by the
time-evolution of the divergence of nearby orbits would then measure
the degree of ergodicity of the mixing process.  In the case of
chaotic mixing, this could lead to a system that does not have much
memory of its evolutionary history. The timescales for evolution could
be relatively short, and in principle, this process could be important
in the path towards equilibrium for galaxies in the Universe
\citep{merritt,vm}.

Since the 1970s N-body simulations have become the standard tool for
studies of the formation and dynamics of structures in the
Universe. The question of whether they are a faithful representation
of the Universe has always attracted significant attention. This is
especially true in recent years \citep{diemand,binney}, particularly
with the finding that dark-matter halos have universal density
profiles \citep{nfw,moore,weinberg-a,weinberg-b}.

One of the first works to focus on how N-body systems evolve was
\citet{miller}. Using what must have been the very first computers in
the world, he simulated a self-consistent system in virial equilibrium
of 8 upto 32 particles distributed randomly in a cubic volume. Miller
found that the trajectories of neighbouring particles initially
diverged exponentially. This initial transient has been confirmed
using numerical experiments with a significantly larger number of
particles \citep{lecar,Kandrup-Smith, vm, hm}, as well as with various
degrees of numerical softening \citep{ks01}. This implies that the
initial exponential divergence cannot be purely attributed to the very
grainy nature of the gravitational potential in Miller's experiments.
Furthermore, even in high-resolution N-body realizations of
well-behaved integrable systems such as the Plummer sphere, nearby
orbits experiment a phase of exponential separation at very early
times \citep{kandrup-sideris}. This initial exponential divergence
present in N-body simulations is now known as ``Miller's
instability''. Understanding this puzzle is the focus of this paper.

That N-body systems would show a certain degree of chaoticity is not
necessarily unexpected. However, it seems natural to expect that the
larger the number of particles used to represent an otherwise
integrable smooth gravitational potential, the more faithful the
representation, and hence the lesser the degree of ``numerical'' chaos
\citep{quinlan-tremaine}. There is now significant evidence that when
such a system is represented by a sufficiently large number of
particles, it does tend to the behaviour expected from the
collisionless Boltzmann equation
\citep{goodman,elzant,kandrup-sideris,sideris}.

Nevertheless, even in these high-resolution experiments the initial
exponential growth phase is present \citep{Kandrup-Smith,vm,ks01}.
Furthermore, there is evidence \citep{goodman,hm} that the rate of
divergence associated to this phase increases in proportion to the
number of particles used. Because Miller's instability only lasts for
a very short timescale this does not imply that the system is
(macroscopically) chaotic \citep{vm}.  As stated by \citet{elzant} it is
likely that the ``mechanism leading to the short e-folding time in
point particle systems is physically unimportant''.

So, while the existence of a continuum limit in N-body systems appears
to be more or less established for long timescales, on short
timescales Miller's instability remains a puzzle. The physical
mechanism responsible for this was hitherto unknown. It seems quite
unlikely that collisions between particles could be important on
timescales as short as one-tenth of the crossing time of the system,
as measured for example by \citet{hm}. Microscopic chaos arising from
``white-noise'' or poor orbit integrations are also unlikely to be
important on those timescales, particularly in integrable
(well-behaved) potentials.

In this paper, we tackle this paradox by studying the {\it initial}
behaviour of nearby characteristics in an integrable smooth (and
analytic) potential. Our aim is to understand how these nearby
characteristics diverge on short timescales, and if they do so at
nearly exponential rates. As we shall demonstrate below, this is
indeed the case. The initial behaviour mimics an exponential
divergence, but since the system is fully integrable this is not
related to the presence of chaos. This near-exponential behaviour
merely reflects the time evolution of an orbit in phase-space. This
result shows that there is no need to introduce the concept of
microscopic chaos, and confirms previous suspicions that this
transient phenomenon is not related to an instability in the sense of
non-integrable behaviour in the dynamics of N-body systems.

In this paper we describe the evolution of nearby orbits in
phase-space, and in particular in configuration space, expanding upon a
model developed by \citet{hw} (hereafter HW). The details of this
formalism are given in Sec.~\ref{sec:aa}. In this Section we focus in
detail on the behaviour of nearby orbits in a Plummer potential. In
Sec.~\ref{sec:comp} we summarize our results.

\section{The evolution in phase-space of nearby orbits in 
integrable potentials} 
\label{sec:aa}

\begin{figure}
\hspace*{-0.5cm}
\includegraphics[width=55mm, angle=270]{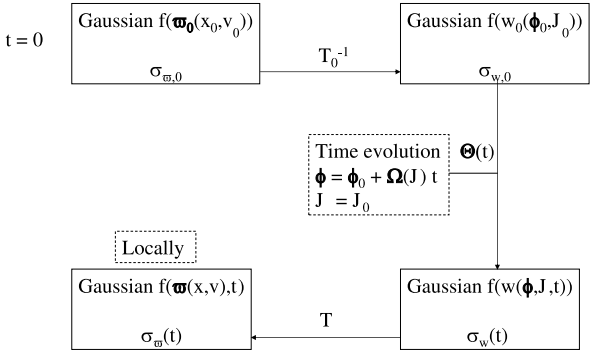}
\caption{Flow chart showing the basic steps of our analytic formalism
to measure the evolution of a system in phase-space.}
\label{fig:chart}
\end{figure}

The problem of the phase-space evolution of nearby orbits has many
applications. Some of the most recent are related to the evolution of
streams formed by the disruption of satellite systems (dwarf galaxies,
globular clusters) in an external (Galactic) potential. This is also
the basis of the formalism that HW developed, which is based on the
conservation of phase-space density. It consists in a mapping from the
initial to the final configurations using adiabatic invariants (a
schematic flow chart is given in Figure \ref{fig:chart}).

The basic idea is to map the initial system onto action-angle space,
then follow the much simpler evolution in this space, and finally
transform back {\it locally} onto observable coordinates (all these
being linear transformations; for details see HW). This method, which
uses action-angle variables, is very general and can be applied to any
potential that admits regular orbits \citep{goldstein,bt}. However, if
the potential is separable, the implementation is simpler. This
includes all spherically symmetric potentials but only few
axisymmetric and triaxial cases, such as the general class of
St\"ackel potentials e.g. \citet{lb62,tim85,tim88}.  In this paper we
shall only focus on spherical potentials because these are the
simplest to model, while at the same time, they evidence a generic
behaviour.

Therefore, instead of following the evolution of pairs of nearby
orbits as is traditional in N-body systems, we follow the evolution of
a distribution function in phase-space. In particular, and for
simplicity, we assume this distribution function to be a multivariate
Gaussian (in 6-dimensions).

The work presented here exploits and expands in two new directions the
HW algorithm. Firstly, we now compute the behaviour of streams in
physical space (to be able to determine the evolution of the spatial
separation of nearby trajectories). Secondly, we derive explicitly new
analytic expressions for this evolution on short timescales.

\subsection{The evolution of the distribution function}

As discussed above, we assume that the initial distribution function
of the system is a multivariate Gaussian in $\varpi = ({\bf x}, {\bf
v})$ coordinates centered on $\langle \varpi_0\rangle$ (a given
particle or orbit):
\begin{equation}
f(\varpi,t_0) = f_0 \exp{\left[-\frac{1}{2} {{\bf
\Delta}^{\dagger}_{\varpi,0} {\bf \sigma}_{\varpi,0} {\bf
\Delta}_{\varpi,0}} \right]}
\end{equation}
where ${\bf \Delta}_{\varpi,0} = \varpi - \langle \varpi_0 \rangle$,
and ${\bf \sigma}_{\varpi,0}$ is the variance matrix (the inverse of
the covariance matrix) at the initial time:
\begin{equation}
\label{eq:matrix_sigma}
{\bf \sigma}_{\varpi,0} = \left[\begin{array}{cc} 
 {\bf{S_{x,0}}} & {\bf C_{xv,0}} \\
 {\bf C_{xv,0}} & {\bf{\sigma_{v,0}}} \end{array}\right].
\end{equation}   
For example, if the variance matrix is diagonal, then ${\bf
S_{x}} = [1/\sigma_{x_i}^{2} \delta_{ij}]$ and ${\bf{\sigma_{v}}}
=[1/\sigma_{v_i}^{2}\delta_{ij}]$, and ${\bf C_{xv}} = {\bf 0}$. 

A mapping ${\bf T} : \varpi \leftarrow w = ({\bf \phi}, {\bf J})$ will
be linear provided the extent of the system in phase-space is
small. Its elements are $T_{ij} = \partial \varpi_i/\partial w_j$
evaluated at $\langle \varpi \rangle$.  Such a mapping will preserve
the form of the distribution function. This will now be a Gaussian in
action-angle space, with variance matrix ${\bf \sigma}_{w,0} = {\bf
T}_0^\dagger {\bf \sigma}_{\varpi,0} {\bf T}_0$.

The dynamical evolution of the system in action-angle coordinates is
given by
\begin{equation}
{\bf \phi} = {\bf \phi}_0 + {\bf \Omega(J)} t, \qquad {\bf J = constant}.
\end{equation}
We may express 
\[{\bf \Delta}_w = {\bf \Theta}(t) {\bf \Delta}_{w,0} \qquad
\mbox{since} \quad
\Delta \phi_{0,i} \sim - \Delta \phi_i - \frac{\partial
\Omega_i}{\partial J_j} \Delta J_j t,\] and where
\begin{equation}
\label{eq:matrix_th}
{\bf \Theta}(t) = \left[\begin{array}{cc} 
 {\bf{\cal I}_3} & - {\bf \Omega'} t \\
 {\bf 0} & {\bf{\cal I}_3} \end{array}\right].
\end{equation}   
${\bf{\cal I}_3}$ here is the identity matrix in 3-D, and ${\bf
\Omega'}$ represents a $3\times3$ matrix whose elements are $\partial
\Omega_i/\partial J_j$.

Therefore the distribution function at time $t$ is
\begin{equation}
\label{eq:ft}
f({\bf w},t) = f_0 \exp{\left[-\frac{1}{2} {{\bf
\Delta}_{w}}^{\dagger} {\bf \sigma}_w {\bf
\Delta}_w \right]},
\end{equation}
where ${\bf \sigma}_w$ is now a function of time
\begin{equation}
{\bf \sigma}_w = {{\bf\Theta}(t)}^{\dagger} {\bf \sigma}_{w,0}{\bf \Theta}(t).
\end{equation}

Finally, using Eq.~(\ref{eq:ft}) we may derive the distribution
function in configuration and velocity space at time $t$. To this end,
we perform a local transformation using the matrix {\bf T}. Since this
is done locally, our distribution function is still a multivariate
Gaussian. The variance matrix at time $t$ is
\begin{equation}
\label{eq:sigma_final}
{\bf \sigma}_{\bf \varpi}(t) = ({\bf T}_0 {\bf \Theta}(t) 
{\bf T}^{-1})^{\dagger} 
{\bf \sigma}_{{\bf \varpi}, 0} ({\bf T}_0 {\bf \Theta}(t) {\bf T}^{-1}).
\end{equation}

The variance matrix contains all the information about the properties
of the particles on initially nearby orbits. For example, the
evolution of the velocity ellipsoid may be derived from the velocity
submatrix: $\sigma_{\bf v}$. This submatrix describes the velocity
distribution of nearby particles at time $t$. The spatial density at a
particular location ${\bf x}$ at time $t$ (which is related to the
spatial separation of those particles) is obtained by integrating the
distribution function with respect to all velocities:
\begin{equation}
\label{eq:rho_xt}
\rho({\bf x},t) = (2 \pi)^3 f_0 \sigma_{v_1} \sigma_{v_2} \sigma_{v_3}
\times \exp {\left[-\frac{1}{2} {{\bf \Delta}_x}^{\dagger} {\bf
\sigma_x}{\bf \Delta}_x \right]}
\end{equation}
where $\sigma_{v_{i=1,2,3}}$ are the velocity dispersions along the
principal components of the velocity ellipsoid. The matrix ${\bf
\sigma_x}$ is 3$\times$3, and contains all the information concerning
the evolution of the particle distribution in configuration space,
including their separation, which is ultimately, the quantity that we
want to measure.

\begin{figure}
\includegraphics[width=82mm]{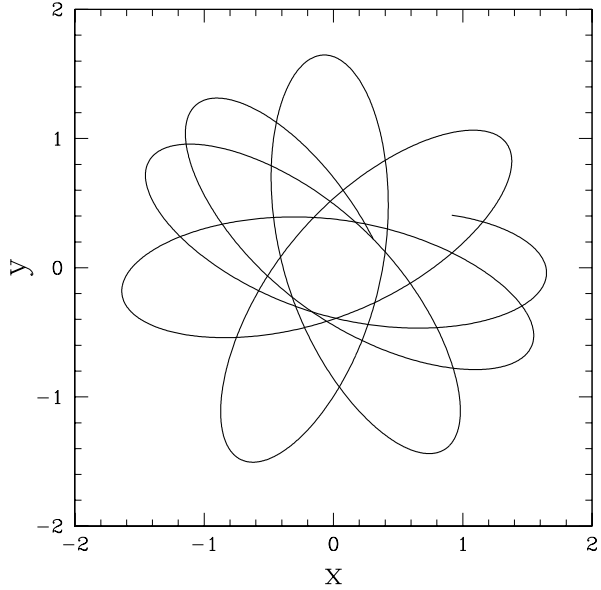}
\caption{Example of an orbit integrated in a Plummer potential.}
\label{fig:orbit}
\end{figure}

\begin{figure*}
\includegraphics[width=83mm]{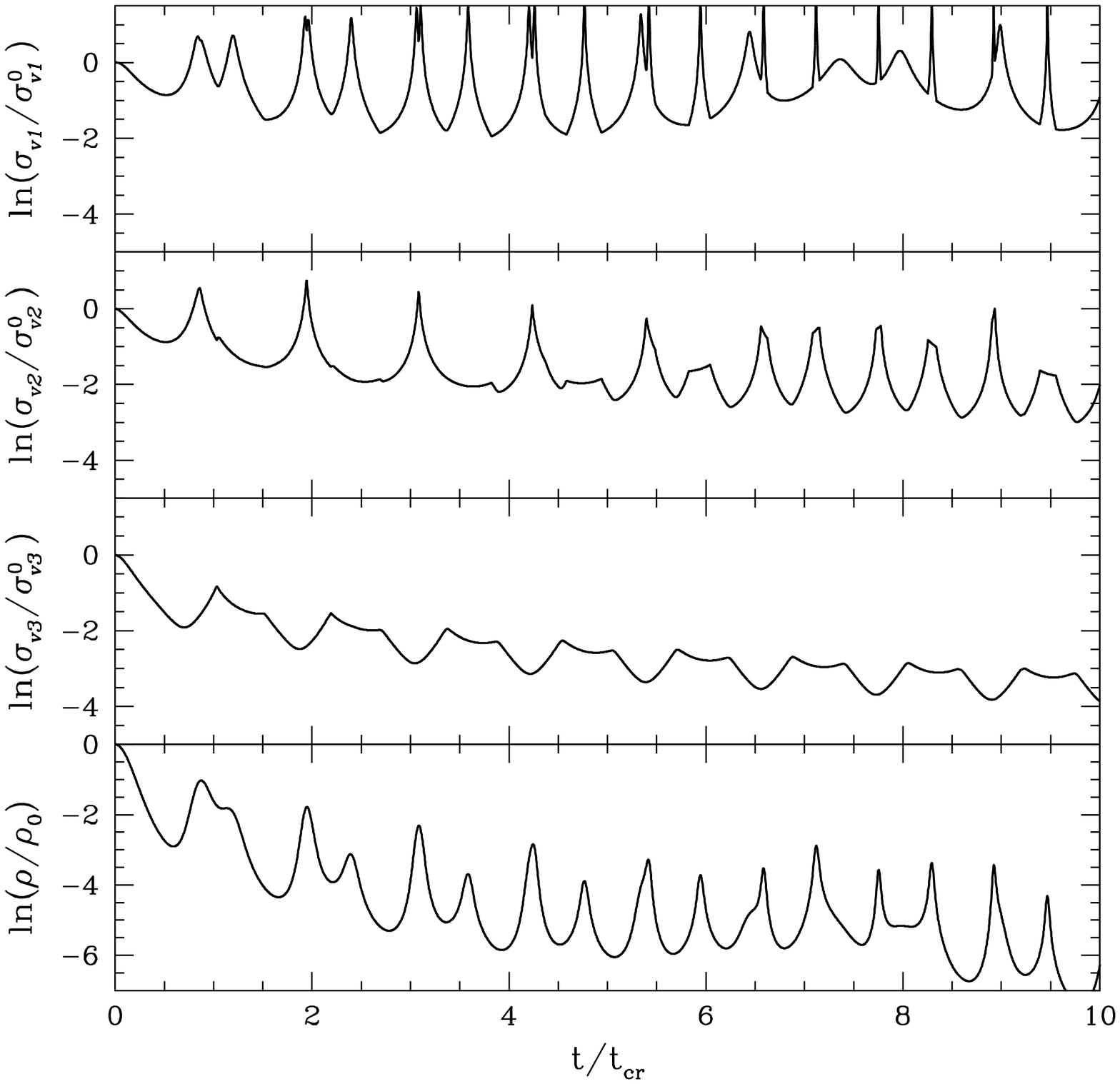}
\includegraphics[width=83mm]{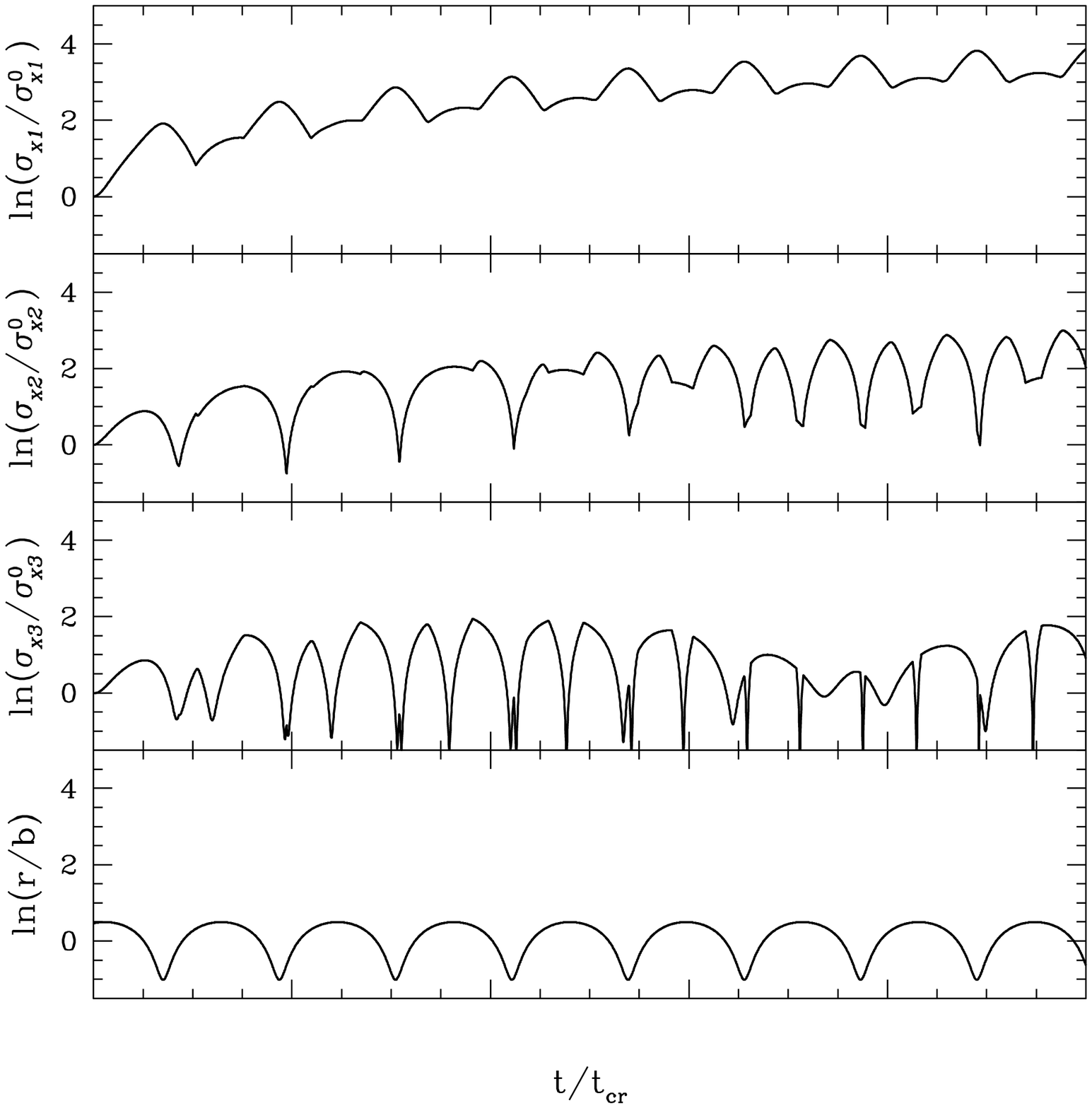}
  \caption{Time evolution of the velocity dispersions (top three
panels on the left), spatial density (bottom left panel) and
dispersions in configuration space (top three panels on the right),
for a system moving in a Plummer potential on the orbit shown in
Figure ~\ref{fig:orbit}. The periodicity observed is related to the
radial (and angular) orbital oscillations, as shown in the bottom
panel on the right.}
\label{fig:ex1}
\end{figure*}

\subsection{Example: Evolution in a Plummer potential}

This simple spherical gravitational potential has the form 
\begin{equation}
\phi(r)=-{\frac{GM}{\sqrt{r^2 + b^2}}}.
\end{equation}
Units were so chosen that $G=M=b=1$ and the internal energy of the
system is $E = -\frac{3\pi}{64}$. We define the crossing time of the
system $t_{cr} = R/V$ where $R=-GM^2/2E$ and $V^2=-2E/M$. For example,
for a dwarf galaxy size system with $b=0.5$ kpc and $M = 10^7
M_\odot$, then $t_{cr} \sim 0.33$ Gyr.

We assume that the initial 6D variance matrix ${\bf
\sigma}_{\varpi}^{0}$ is diagonal (see Eq.~\ref{eq:matrix_sigma}),
with ${\bf S_{x,0}} = [1/\sigma_x^{2} \delta_{ij}]$ and
${\bf{\sigma_{v,0}}} = [1/\sigma_v^{2} \delta_{ij}]$, and where $\sigma_x
= 10^{-5}$ and $\sigma_v = 10^{-5}$. We set the central particle
of the system on an orbit whose apocentre is located at $r_a=1.635 b$,
as shown in Fig.~\ref{fig:orbit}.

In Figure~\ref{fig:ex1} we plot the evolution of the velocity
dispersions, the spatial density and the dispersions in configuration
space as function of time, for the orbit shown in
Figure~\ref{fig:orbit}. These are computed using the procedure outlined
in the previous section.

Figure \ref{fig:ex1} shows that in the case of spherical potentials,
only two of the velocity dispersions decrease in time, while the third
one remains on average constant (it corresponds to the direction
perpendicular to the plane of motion). These results imply that the
configuration-space dispersions will increase in time, as a consequence
of Liouville's theorem (i.e. the conservation of phase-space
density). This can also be seen from $dM \sim \rho \times \sigma_{x_1}
\sigma_{x_2} \sigma_{x_3} = cst$.

The form of the dispersions in velocity and in configuration space has
been derived explicitly in the Appendix. There we work in a reference
frame that coincides with the plane of motion (this is of course
possible for a spherical potential). In this new frame only two
coordinates and two velocities are required to specify completely the
state of system. In this case, the spatial density
\begin{equation}
\label{eq:rho_sigma}
\rho \propto \sigma_{v_{1}} \sigma_{v_{2}} =
(\lambda_{v_{1}}\lambda_{v_{2}})^{-1/2},
\end{equation}
 where $\lambda_v$ denotes the eigenvalues of the velocity submatrix
 ${\bf \sigma_v}$, and for which the following relation holds
\begin{equation}
\label{eq:sigma_vel_ev_maintext}
\lambda_{v_{1}}\lambda_{v_{2}}=\frac{r^{2}p_{r}^{2}}{\Omega_{r}^{2}}\left(\alpha_{4}t^{4}+\alpha_{3}t^{3}+\alpha_{2}t^{2}+\alpha_{1}t+\alpha_{0}\right).
\end{equation}
The coefficients $\alpha_i$ depend both on location along the orbit as
well as on the initial extent of the system in phase-space (see
Eq.~\ref{eq:alphas}). The very rapid decrease in the spatial density
of the system observed in Fig.~\ref{fig:ex1} can be understood from
Eqs.~(\ref{eq:rho_sigma}) and (\ref{eq:sigma_vel_ev_maintext}). This
decrease implies a rapid separation of the particles (and hence of
their orbits). Furthermore, the strong enhancements in the density
seen in Fig.~\ref{fig:ex1} take place at the orbital turning points:
when $p_r = 0$ then $\lambda_{v_1} \times \lambda_{v_2} \rightarrow
0$ and hence $\rho \rightarrow \infty$.

In the Appendix (see Eq.~\ref{eq:lambda_r}), we show that the
configuration-space dispersions $\sigma_{x_1} \sigma_{x_2} =
\D{\sqrt{\frac{\lambda_{v_{1}}\lambda_{v_{2}}}{\det{{\bf
\sigma}_{w}^0}}}}$. Close inspection of
Eqs.~(\ref{eq:sigma_vel_ev_maintext}) and (\ref{eq:alphas}), allows us to reach
the following conclusions:
\begin{itemize}
\item For very short timescales, the term with $\alpha_0$
dominates. In this case the separation of nearby orbits as measured by
$\sigma_x$ purely reflects the geometry of the orbit in phase space
(being heavily weighted by $r^2 p_r^2$).
\item The terms with $\alpha_2$ and $\alpha_4$ are always positive,
implying that these will induce a rapid increase in the $\lambda_v$,
and hence of the dispersions in configuration space on intermediate
timescales.
\item The terms with $\alpha_1$ and $\alpha_3$ can either be positive
or negative, depending on location along the orbit. This (partly)
explains the strong oscillatory behaviour observed in
Fig.~\ref{fig:ex1}.
\item On longer timescales, only the term $\alpha_4 t^4$ is important.
This gives rise to the secular behaviour of density which decreases as
$1/t^2$ (as found by HW), and for the dispersions in
configuration-space to increase as $t$.
\end{itemize}

\subsubsection{Relating the dispersion to the separation in 
configuration space}
\label{sec:means}

Our main aim is to study the separation of initially nearby orbits in
configuration space. The question is how this separation is related to
the dispersions or the variances (i.e. the inverse of the eigenvalues)
in the configuration-space matrix $\sigma_{\bf x}$.

We examine three possibilities obtained by performing three different
kinds of averages of the configuration-space dispersions:
\begin{enumerate}
\item the geometric mean: $\Delta_{g} = (\sigma_{x_1} \sigma_{x_2}
\sigma_{x_3})^{1/3}$,
\item the arithmetic mean: $\Delta_{a}
=(\sigma_{x_1}+\sigma_{x_2}+\sigma_{x_3})/3$,
\item the modulus: $\Delta_{m} =
\sqrt{\sigma_{x_1}^2+\sigma_{x_2}^2+\sigma_{x_3}^2}/3$.
\end{enumerate}

\begin{figure}
\includegraphics[width=83mm]{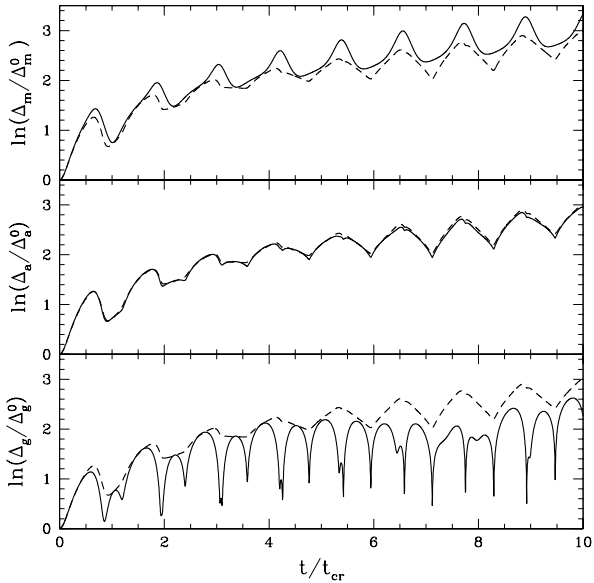}
  \caption{Time evolution of three possible averages of the
dispersions in configuration-space, as obtained through our
formalism. Note the very fast initial growth and the linear behaviour
at late times. The dashed curve in each panel represents the average
separation $\langle \Delta_r \rangle$ of 1000 nearby orbits. Note the
excellent agreement between $\langle \Delta_r \rangle$ and
$\Delta_a$ (middle panel).}
\label{fig:aver}
\end{figure}

Figure \ref{fig:aver} shows the behaviour of these different
averages. In all cases, one observes initially a very fast increase in
the measured values, while at late times the growth proceeds
linearly with time [see also Eqs.~(\ref{eq:sigma_vel_ev}) and
(\ref{eq:sigma_r_ev})].

The question now is how to relate the above defined averages to the
separations between two initially nearby orbits. To address this we
have generated 1000 orbits with initial conditions distributed
according to a Gaussian in configuration space with dispersion
$\sigma_x = 10^{-5}$ and $\sigma_v = 10^{-5}$ around the orbit shown
in Fig.~\ref{fig:orbit}. We measure the separation $\Delta_r = |{\bf
r_i} - {\bf r_0}|$ between this orbit and the 1000 neighbouring
trajectories, and derive the average $\langle \Delta_r \rangle$. This
is plotted as a dashed curve in the panels of Fig.~\ref{fig:aver}. As
can be seen, the arithmetic mean $\Delta_a$ of the configuration-space
dispersions computed using the analytic formalism discussed in the
previous section provides an excellent measurement of the separation
between nearby orbits.

Figure \ref{fig:aver} shows that the separation of nearby orbits in
smooth integrable potentials exhibits an initial rapid divergence,
which is followed by a secular increase which is linear in time. Note
that this occurs for a completely integrable system, without any
degree of chaos. Therefore we see already that the initial exponential
divergence cannot be attributed to any form of chaos whatsoever.  It
simply reflects the way an orbit evolves in phase-space, as shown in
the Appendix.

\subsubsection{Direct comparison to N-body simulations}

The behaviour visible in Figure \ref{fig:aver} is strinkingly similar
to that observed in the N-body simulations shown in Fig.~1 of
\citet{vm} and in the frozen N-body realizations of
\citet{kandrup-sideris}. To provide the reader with a more direct
comparison to our analytic results, we will now analyse the divergence
of an ensemble of initially nearby orbits evolved in a potential
represented by a finite number of particles.

We have generated a realization of the Plummer sphere, whose density
profile is
\begin{equation}
\rho(r) = \frac{ 3M }{ 4{\pi}b^3}\left( 1 +
\frac{ r^2 }{ b^2 }\right)^{ -5/2 },
\end{equation}
which we have truncated at a radius $r_{t} = 12.197b$, that encloses 99\%
of its mass. We represent this system with $N = 128,000$ particles,
and use a numerical softening ${\epsilon} = 0.025b$ (see
e.g. \cite{atha}). Following \cite{ks01} we integrate orbits in this
frozen (in time and space) N-body realization. The integration of the
orbits was performed using a Runge-Kutta-Fehlberg algorithm of order
4--5.

As in the previous section, we follow the evolution of 100 orbits
distributed according to a Gaussian in phase space with initial
dispersion ${\sigma}_{x} = 10^{-5}$ and ${\sigma}_{v} = 10^{-5}$
around the orbit shown in Fig. 2. For this ensemble we have measured
the time evolution of the average separation $\langle \Delta_r
\rangle$ and of the 6D variance matrix.

\begin{figure}
\includegraphics[width=83mm]{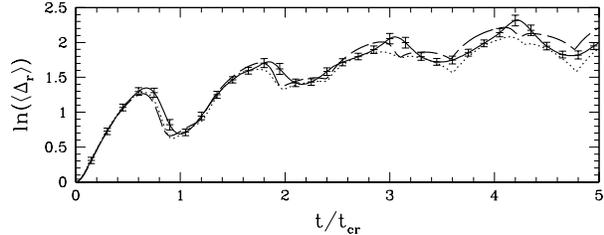}
  \caption{Time evolution of the average separation $\langle \Delta_r
  \rangle$ of 100 nearby orbits integrated in a frozen N-body
  realization of the Plummer sphere (solid curve). The error bars
  correspond to the error on this average. The dashed curve represents
  the separation of nearby orbits as measured by $\Delta_a$ using our
  analytic prescription, while the dotted line represents the same
  quantity but estimated from the 100 orbits integrated in the N-body
  realization.}
\label{fig:aver_num}
\end{figure}

The results are shown in Fig. \ref{fig:aver_num}. The initial
behaviour is very similar, whether derived using our analytic
formalism ($\Delta_a$, dashed curve) or using the average separation
of 100 orbits in the N-body representation of the system ($\langle
\Delta_r \rangle$, solid curve). The small differences can be
attributed to two causes. First of all, the finite sampling of
phase-space around the central orbit introduces an error in the
average, which is quantified by the error bars shown in this
figure. Secondly, $\langle \Delta_r \rangle$ is not exactly identical
to $\Delta_a$ (see Figure \ref{fig:aver}). If we compare the quantity
$\Delta_a$ obtained using the 6D variance matrix from the frozen
N-body simulation (dotted curve) with that from the analytic formalism
(dashed curve), this difference almost disappears. The two curves are
virtually indistiguishable over a timescale of a few crossing times.

\subsubsection{Dependence on initial conditions}

It is also interesting to understand how the separation of initially
nearby orbits depends on their initial conditions. In particular, how
the initial divergence depends on the region of phase-space sampled at
short times.

\begin{figure}
\includegraphics[width=83mm]{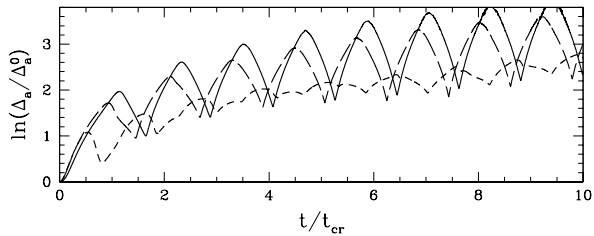}
\caption{Time evolution of the separation of nearby orbits as measured
by the arithmetic mean $\Delta_a$ for the same orbit discussed in
previous figures. The sensitivity to initial location along a given
orbit is evidenced by the various curves: solid corresponds to initial
location at pericentre; short-dashed to apocentre and long-dashed to
the average distance between these turning points.}
\label{fig:apo-peri}
\end{figure}

In Figure \ref{fig:apo-peri} we plot the time evolution of the
arithmetic mean of the three dispersions in configuration space
$\Delta_a$ obtained using our analytic prescription for the orbit
shown in Fig.~\ref{fig:orbit}. We now plot the behaviour for different
starting points along this orbit: apocentre, pericentre and (apocentre
+ pericentre)/2.  We see clearly that the behaviour at short times
depends on the initial location along the orbit. The initial
divergence is in all cases nearly exponential, but has largest
amplitude (it lasts longer) when the integration is started near
pericentre.

It is also interesting to study the behaviour of different sets of
nearby orbits. Figure \ref{fig:fits} shows the evolution for two
additional examples. The top panel corresponds to an orbit constrained
to move in the inner regions of the system (pericentre $r_p = 0.62b$
and apocentre $r_a = 0.88b$), while the bottom panel has $r_p = 1.73
b$ and $r_a = 2.94 b$, and hence it is constrained to the
outskirts. Clearly the amplitude of the initial growth phase depends
on the regions of phase-space the orbits probe.

Note that, because the quantities shown are normalized to their
initial conditions, these results are independent of the initial
separation of the orbits (or the values of $\sigma_{ii}$ in our
formalism). This is perhaps, the most characteristic difference
between an integrable and a chaotic system. The amplitude (or the
rate) of the initial divergence {\it for a given orbit} is always the same
in the integrable case, irrespective of initial separation. To the
contrary, in a chaotic system, in the limit of infinitesimal
perturbations, the orbits may be trapped near a resonance and cease to
be chaotic to become regular. Hence the amplitude of the initial
divergence will, in the chaotic case, depend strongly on the initial
separation of the orbits.

\subsubsection{Miller's instability and the initial behaviour}
\label{sec:lyap}

The above analysis shows that the initial very rapid divergence of
nearby orbits is a generic feature of dynamical systems.  It is not
only observed in the Plummer potential discussed here, but also in all
integrable potentials studied by HW (e.g.\ Fig. 7 and 9 of their
paper).

\begin{figure}
\includegraphics[width=83mm]{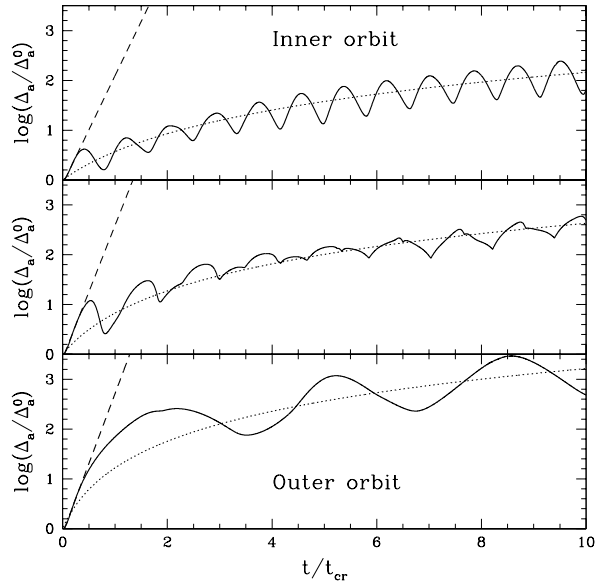}
\caption{Time evolution of the separation of nearby orbits as measured
by the arithmetic mean $\Delta_a$. The various panels represent orbits
probing different regions of the system, all integrated from their
apocentres. The dotted curves are exponential fits to the initial
(transient) behaviour, while the dashed curves are linear fits to
the long-term (secular) behaviour of $\Delta_a$.}
\label{fig:fits}
\end{figure}

The initial behaviour is nearly exponential, as shown in
Fig.~\ref{fig:fits} for the orbits discussed so far. The rate of
divergence --the equivalent of the ``short-term'' Lyapunov exponent,
is $\chi_e \sim 13.6/t_{cr}$ for the inner orbit, $\chi_e \sim
16.7/t_{cr}$ and $\chi_e \sim 17.57/t_{cr}$ for the intermediate and
outer orbits, respectively. In all cases, the secular behaviour at
late times can be fit by a linear function of time $\Delta_a =
\Delta_{a,0} + t/t_{sec}$, where the divergence timescale is $t_{sec}
\sim 1.3 t_{cr}$, $t_{sec} \sim 0.77 t_{cr}$ and $t_{sec} \sim 0.41
t_{cr}$ for the different orbits.

\section{Discussion}
\label{sec:comp}

Our analysis shows that the initial nearly-exponential divergence of
nearby orbits in N-body systems is not due to chaos. It is present
also in integrable smooth potentials, and it reflects a power-law
divergence modulated by the shape of an orbit in phase-space.

It is interesting to note that the rates of divergence that we measure
using our formalism are in very good agreement with those obtained by
\citet{hm} for N-body realizations of the same Plummer sphere. These
authors find a characteristic e-folding time of $t_{cr}/20$ for
systems with $N \sim 10^5$ particles, which is very comparable to the
values obtained in Section \ref{sec:lyap}. They also find a weak
dependence on $N$, which may also be readily understood within our
framework. Such a dependence is induced by the very rapid decrease in
the spatial density of the system.  If a ``relatively'' small number
of particles is used in a N-body simulation, then the density cannot
be mapped properly. For example, to measure a decline in the density
of $10^{-5}$ on a timescale of $\sim 3 t_{cr}$ as observed in
Fig.~\ref{fig:ex1}, N-body realizations with at least
$10^{5}$ nearby particles are needed.

Previous works, including \citet{miller} and \citet{kandrup-sideris}
have also noted an oscillatory behaviour in the divergence of nearby
orbits.  Our analysis, as well as Figure \ref{fig:ex1} show that this
is due to the modulation produced by the periodicity of a regular
orbit in phase-space. It is not, as suggested by \citet{miller}, due
to the formation of tight binaries in an N-body system. The fact that
such behaviour was visible in the various N-body studies presented in
the literature, in fact demonstrates that such N-body systems were
faithful representations of the true (integrable) system, at least on
short timescales. As stated by \citet{Kandrup} and \citet{vm}, the
Lyapunov exponents need to be measured in the limit of infinite time
intervals; short-time exponential-like divergences do not imply
chaotic behaviour.

\section*{Acknowledgments}
We thank Daniel Carpintero for bringing up this problem to our
attention as well as Simon White and Ortwin Gerhard for enlightening
discussions. NWO, NOVA and the Kapteyn Institute are gratefully
acknowledged for financial support.

\appendix

\onecolumn
\section{Matrices in a spherical potential}

For spherical potentials $\Phi(r)$, we may choose a system of
coordinates that coincides with the plane of motion of the system. In
this plane the position of a particle is specified by its angular
$(\psi)$ and radial $(r)$ coordinates. The actions of an orbit in this
case are:
\begin{equation}
L=J_{\psi}=p_{\psi},\hspace{0.5cm}
J_{r}=\frac{1}{\pi}\int_{r_{1}}^{r_{2}}dr\frac{1}{r}\sqrt{2[E-\Phi(r)]r^{2}-L^{2}},
\end{equation} 
where $L$ is the total angular momentum of the particle, $E$ is its
energy, and $r_{1}$ and $r_{2}$ the orbital turning points.

In order to track the evolution of the dispersions of our initial
distribution function, $f(\varpi,t_0)$, we perform the following
sequence of operations.  Firstly, we transform from Cartesian
coordinates ${\bf \varpi}=(\bf{x},\bf{v})$ to action angle variables
$\bf{\it{w}}=(\bf{\theta},\bf{J})$. The distribution function is then
evolved in this space, after which, we transform back to Cartesian
coordinates (see Figure \ref{fig:chart}).

For the sake of simplicity, here we begin with a distribution function
already expressed in terms of the action-angle variables and we also
assume that, initially, the variance matrix is diagonal, i.e.,
$\sigma_{w,0}=[\sigma_{ii}\delta_{ij}]$. With the time evolution
operator, ${{\bf\Theta}(t)}$, known, we can compute the variance
matrix at any given time t as ${\bf \sigma}_{w}(t) =
{{\bf\Theta}(t)}^{\dagger} {\bf \sigma}_{w,0}{\bf
\Theta}(t)$. Equation (\ref{eq:matrix_th}) shows ${{\bf\Theta}(t)}$
for the 3-D case, but we reduce the equation for our purposes to the
2-D case.

After evolving the system in the action-angle space we need to
transform back locally to configuration and momenta space ${\hat {\bf
\omega}}=({\bf x},{\bf p})$ using the transformation matrix
$\bf{T^{-1}}$. The elements of this matrix are related to the second
derivatives of the characteristic function $W(\bf{q},\bf{J})$. In our case 
\begin{equation}
{\bf T^{-1}}=\left[ \begin{array}{cccc}
1 & t_{12} & t_{13} & t_{14} \\
0 & t_{22} & t_{23} & t_{24} \\
0 & 0 & 1 & 0 \\
0 & t_{42} & t_{43} & t_{44} \\
\end{array} \right],
\end{equation}
with\\
\begin{displaymath}
\begin{array}{lll}
t_{12}=\D{-\frac{h(r)}{\Omega_{r}}W_{34}+ \frac{\kappa}{p_{r}}}, &
\D{t_{13}=W_{33}+W_{34}t_{43}}, & \D{t_{14}=W_{34}t_{44}}, \\
t_{22}=\D{-\frac{h(r)}{\Omega_{r}}W_{44}+ \frac{\Omega_{r}}{p_{r}}}, &
t_{23}=\D{W_{34}+W_{44}t_{43}}, & t_{24}=\D{W_{44}t_{44}}, \\
t_{42}=\D{{-\frac{h(r)}{\Omega_{r}}}}, &
t_{43}=\D{-\frac{\kappa}{\Omega_{r}}}, &
t_{44}=\D{\frac{p_{r}}{\Omega_{r}}},
\end{array}
\end{displaymath}
 where \\
\begin{displaymath}
\begin{array}{lll}
\D{h(r)=-\Phi^{\prime}(r)+\frac{L^{2}}{r^{3}}}, &
\D{p_{r}=\sqrt{2[E-\Phi(r)]-\frac{L^{2}}{r^{2}}}}, &
\D{\kappa=\Omega_{\psi}-\frac{L}{r^2}},
\end{array}
\end{displaymath}
and\\
\begin{displaymath}
\begin{array}{l}
\D{W_{33}=\frac{\partial^{2}W}{\partial{L}^{2}}=\int_{r_1}^{r}\frac{dr}{p_r}\left(\frac{\partial{\Omega_{\psi}}}{\partial{J_{\psi}}}-\frac{1}{r^2}-\frac{\kappa^{2}}{p_{r}^{2}}\right)},\\\\
\D{W_{44}=\frac{\partial^{2}W}{\partial{J_{r}}^{2}}=\int_{r_1}^{r}\frac{dr}{p_{r}}\left(\frac{\partial{\Omega_{r}}}{\partial{J_{r}}}-\frac{\Omega_{r}^{2}}{p_{r}^{2}}\right),}\\\\
\D{W_{34}=\frac{\partial^{2}W}{\partial{L}\partial{J_{r}}}=\int_{r_1}^{r}\frac{dr}{p_{r}}\left(\frac{\partial{\Omega_{\psi}}}{\partial{J_{r}}}-\frac{\kappa}{p_{r}^{2}}\Omega_{r}\right).}\\\\
\end{array}
\end{displaymath}

Subindices {1} and {3} in the expressions above refer to directions
associated with $\psi$, such as, $\phi_{\psi}$ and $J_{\psi}$ whereas
{2} and {4} are related to $r$. For more details about this procedure
we refer the reader to HW.

Given ${\bf T^{-1}}$, the variance matrix at time $t$ is expressed as:
\begin{equation}
\label{eq:T}
{\bf \sigma}_{\hat{\bf \omega}}(t) = ({\bf \Theta}(t) {\bf T}^{-1})^{\dagger}
{\bf \sigma}_{{\bf \it{w}}}^{0}({\bf \Theta}(t) {\bf T}^{-1}).
\end{equation}
where the elements $t_{ij}$ are evaluated at $\langle {\bf x}(t)
\rangle$. Substituting $\bf{T^{-1}}$, ${{\bf\Theta}(t)}$ and ${\bf
\sigma}_{{\bf \it{w}}}^{0}$ in the above expression, then 
\begin{equation}
{\bf \sigma_{\hat{\omega}}}(t)=\left[ \begin{array}{cccc} \sigma_{11} &
\sigma_{11}A & \sigma_{11}B & \sigma_{11}C \\ \{1,2\} &
\sigma_{11}A^{2}+\sigma_{22}D^{2}+\sigma_{44}t_{42}^{2} &
\sigma_{11}AB+\sigma_{22}DE+\sigma_{44}t_{42}t_{43} &
\sigma_{11}AC+\sigma_{22}DF+\sigma_{44}t_{42}t_{44} \\ \{1,3\} &
\{2,3\} &
\sigma_{11}B^{2}+\sigma_{22}E^{2}+\sigma_{33}+\sigma_{44}t_{43}^{2} &
\sigma_{11}BC+\sigma_{22}EF+\sigma_{44}t_{43}t_{44} \\ \{1,4\} &
\{2,4\} & \{3,4\} &
\sigma_{11}C^{2}+\sigma_{22}F^{2}+\sigma_{44}t_{44}^{2} \\
\end{array} \right],
\end{equation}
where\\ \\
$A=t_{12}-{\Omega^{\prime}_{34}}t_{42}t$,\hspace{0.4cm}$B=t_{13}-({\Omega^{\prime}_{33}}-{\Omega^{\prime}_{34}}t_{43})t$,\hspace{0.5cm}
$C=t_{14}-{\Omega^{\prime}_{34}}t_{44}t$,\\
$D=t_{22}-{\Omega^{\prime}_{44}}t_{42}t$,\hspace{0.4cm}$E=t_{23}-({\Omega^{\prime}_{34}}-{\Omega^{\prime}_{44}}t_{43})t$,\hspace{0.5cm}
$F=t_{24}-{\Omega^{\prime}_{44}}t_{44}t$.\\

In general, one is more interested in the properties of the debris in
velocity space, rather than in momenta space. Therefore we transform the
variance matrix according to ${\bf \sigma_{\varpi}}(t)= {\bf T}_{p
\rightarrow v}^{\dagger}{\bf \sigma}_{\hat{\bf \omega}}(t){\bf T}_{p
\rightarrow v}$, with

\begin{displaymath}
{\bf T}_{p \rightarrow v}=\left[\begin{array}{cccc} 1 & 0 & 0 & 0 \\ 0
& 1 & 0 & 0 \\ 0 & v_{\psi} & r & 0 \\ 0 & 0 & 0 & 1
\end{array}
\right].
\end{displaymath}

To obtain an expression for the time evolution of the velocity
dispersions we focus our attention on what happens around a particular
point $\langle {\bf x}(t) \rangle$ in configuration space located on
the mean orbit of the system. This is equivalent to studying the
velocity submatrix of the variance matrix ${\bf \sigma_{\varpi}}(t)$, that is
\begin{equation}
{\bf \sigma}_{\bf v} = \left[\begin{array}{cc}
 {r^{2}(\sigma_{11}B^{2}+\sigma_{22}E^{2}+\sigma_{33}+\sigma_{44}t_{43}^{2})}
 & {r(\sigma_{11}BC+\sigma_{22}EF+\sigma_{44}t_{43})} \\
 \left\{1,2\right\} &
 {\sigma_{11}C^{2}+\sigma_{22}F^{2}+\sigma_{44}t_{44}^{2}}
 \end{array}\right].
\end{equation}   

By diagonalizing the matrix ${\bf \sigma}_{\bf v}$ we obtain the
principal axes of the velocity ellipsoid at the point $\langle {\bf
x}(t) \rangle$, and the associated dispersions. The eigenvalues of
${\bf \sigma}_{\bf v}$ are the roots of the characteristic equation:
$\det[{\bf \sigma}_{\bf v}-\lambda {\bf \cal I}]=0$. An interesting
quantity is for example, $\lambda_{v_{1}}\lambda_{v_{2}}$ because it
is inversely proportional to the density: $\rho \propto \sigma_{v_{1}}
\sigma_{v_{2}} = (\lambda_{v_{1}}\lambda_{v_{2}})^{-1/2}$. In our case:

\begin{equation}
\label{eq:sigma_vel_ev}
\lambda_{v_{1}}\lambda_{v_{2}}=\frac{r^{2}p_{r}^{2}}{\Omega_{r}^{2}}\left(\alpha_{4}t^{4}+\alpha_{3}t^{3}+\alpha_{2}t^{2}+\alpha_{1}t+\alpha_{0}\right),
\end{equation}
where
\begin{eqnarray}
\label{eq:alphas}
\alpha_{4} & = & \D{\sigma_{11}\sigma_{22}(\det{\bf \Omega^{\prime}})^{2}},  \nonumber \\
\alpha_{3} & = & \D{2\sigma_{11}\sigma_{22}\det{\bf \Omega^{\prime}}\left(2W_{34}{\Omega^{\prime}_{34}}-W_{33}{\Omega^{\prime}_{44}}-W_{44}{\Omega^{\prime}_{33}}\right)}, \nonumber \\
\alpha_{2} & = & \D{\sigma_{11}\sigma_{22}\big(2\det{\bf \Omega^{\prime}}\det{\bf W}+({\Omega^{\prime}_{44}}W_{33}+{\Omega^{\prime}_{33}}W_{44})^{2}+4W_{34}({{\Omega^{\prime}}^2_{34}}W_{34}-{\Omega^{\prime}_{33}}{\Omega^{\prime}_{34}}W_{44}-{\Omega^{\prime}_{34}}{\Omega^{\prime}_{44}}W_{33})\big)+} \nonumber \\
& & \D{\left(\sigma_{11}\sigma_{33}+\sigma_{22}\sigma_{44}\right){{\Omega^{\prime}}^2_{34}}+\sigma_{11}\sigma_{44}{{\Omega^{\prime}}^2_{33}}+\sigma_{22}\sigma_{33}{{\Omega^{\prime}}^2_{44}}}, \nonumber \\
\alpha_{1} & = & \D{2\big(\sigma_{11}\sigma_{22}\det{\bf W}\left(2{\Omega^{\prime}_{34}}W_{34}-{\Omega^{\prime}_{44}}W_{33}-{\Omega^{\prime}_{33}}W_{44}\right)-{\Omega^{\prime}_{34}}W_{34}\left(\sigma_{11}\sigma_{33}+\sigma_{22}\sigma_{44}\right)-\sigma_{11}\sigma_{44}{\Omega^{\prime}_{33}}W_{33}-} \big . \nonumber \\
& & \big .\D{\sigma_{22}\sigma_{33}{\Omega^{\prime}_{44}}W_{44}\big)}, \nonumber \\
\alpha_{0} & = & (\sigma_{11}\sigma_{22})(\det{\bf W})^{2}+ W_{34}^{2}\left(\sigma_{11}\sigma_{33}+\sigma_{22}\sigma_{44}\right)+\sigma_{11}\sigma_{44}W_{33}^{2}+\sigma_{22}\sigma_{33}W_{44}^{2}+\sigma_{33}\sigma_{44},
\end{eqnarray}
\\ with\\ \\ $\det{\bf W}=W_{33}W_{44}-W_{34}^{2}$.\\ \\
These equations explicitly show the behaviour of principal axes velocity
dispersions:
\begin{itemize}
\item For very short timescales, the term with $\alpha_0$
dominates. In this case the behaviour purely reflects the geometry of
the orbit in phase space (being heavily weighted by $r^2 p_r^2$).
\item The terms with $\alpha_2$ and $\alpha_4$ are always positive,
implying that these will induce a rapid increase in the $\lambda_v$,
or a rapid decrease of the velocity dispersions on intermediate timescales.
\item The terms with $\alpha_1$ and $\alpha_3$ can either be positive
or negative, depending on location along the orbit (i.e. the $W_{ij}$
vary in magnitude and sign). This explains the strong
oscillatory behaviour observed in Fig.~\ref{fig:ex1}.
\item On longer timescales, only the term $\alpha_4 t^4$ is important.
This gives rise to the secular behaviour of density which decreases as
$1/t^2$, and the velocity dispersions to behave as $1/t$ for long
timescales.
\end{itemize}

To obtain the expression for the time evolution of the dispersions in
configuration space we integrate the distribution function with
respect to all velocities (see Eq.~\ref{eq:rho_xt}).  In practice, we
first transform ${\bf \sigma}_{\bf \varpi}(t)$ from polar to Cartesian
coordinates, ${\bf \sigma}_{\bf \varpi}^{\prime}(t)=({\bf
T}^{\prime})^{\dagger}{\bf \sigma}_ {{\bf \varpi}}(t){\bf
T}^{\prime}$, where\\

\begin{equation}
{\bf T}^{\prime}=\left[\begin{array}{cccc}
 -\displaystyle \frac{\sin(\psi)}{r} & 
\displaystyle \frac{\cos(\psi)}{r} & 0 & 0 \\
 \cos(\psi) & \sin(\psi) & 0 & 0 \\
\displaystyle \frac{\sin(\psi)p_{r}}{r} & 
\displaystyle -\frac{\cos(\psi) p_{r}}{r} & -\sin(\psi) & \cos(\psi) \\
 \displaystyle -\frac{\sin(\psi)v_{\psi}}{r} & 
\displaystyle \frac{\cos(\psi) v_{\psi}}{r} & \cos(\psi) & \sin(\psi)
\end{array} \right].
\end{equation}
\vspace{0.5cm}
We express ${\bf \sigma}_{\varpi}^{\prime}$ as
\begin{displaymath}
{\bf \sigma}_{\varpi}^{\prime}=\left(
\begin{array}{cc}
{\bf A} & {\bf B}\\
{\bf B}^{\dagger} & {\bf C}\\
\end{array}
\right),
\end{displaymath}
where the 2x2 matrices ${\bf A}$, ${\bf C}$ and ${\bf B}$ represent
the position submatrix, the velocity submatrix, and the cross
correlation between positions and velocities, respectively (as in
Eq.~\ref{eq:matrix_sigma}). Then, the matrix ${\bf \sigma_x}$ is obtained
from the integration of the distribution function over the velocities:
\begin{displaymath}
{\bf \sigma_x}=\left(
\begin{array}{cc}
s_{11} & s_{12}\\
s_{12} & s_{22}\\
\end{array}
\right),
\end{displaymath}
where the elements $s_{ij}$ are related to the dispersions in
configuration space. These elements can be expressed as:
\begin{equation}
s_{ij}=\frac{\det{\bf \Gamma}_{ij}}{\det{\bf C}},
\end{equation}
with
\begin{displaymath}
{\bf \Gamma}_{ij}=\left(
\begin{array}{ccc}
a_{ij} & b_{i1} & b_{i2}\\
b_{j1} & c_{11} & c_{12}\\
b_{j2} & c_{12} & c_{22}\\
\end{array}
\right),
\end{displaymath}
where $a_{ij}$, $b_{ij}$ and $c_{ij}$ are elements of the matrices
${\bf A}$, ${\bf B}$ and ${\bf C}$ respectively. The diagonalization
of the matrix ${\bf \sigma_x}$ yields the values of the dispersions along the
principal axes of the system in configuration space since $\sigma_{x_i} = 1/\sqrt{\lambda_{r_i}}$,
where $\lambda_{r_i}$ are the eigenvalues of ${\bf \sigma_x}$. 

Solving the characteristic equation for ${\bf \sigma_x}$ we finally obtain:
\begin{eqnarray}
\label{eq:sigma_r_ev}
\lambda_{r_{i}} & = & (2\lambda_{v_{1}}\lambda_{v_{2}})^{-1}\left[\beta_{2}t^{2}+\beta_{1}t+\beta_{0}\pm\sqrt{(\beta_{2}t^{2}+\beta_{1}t+\beta_{0})^{2}-4\lambda_{v_{1}}\lambda_{v_{2}}\det{{\bf \sigma}_{w}^0}}\right],
\end{eqnarray}
where
\begin{eqnarray}
\beta_{2} & = &
\D{\sigma_{11}\sigma_{22}r^2\big[(\sigma_{44}{{\Omega^{\prime}}^2_{34}}+\sigma_{33}{{\Omega^{\prime}}^2_{44}})
(p_{r}^{2}+r^{2}\kappa^{2})
-2{\Omega^{\prime}_{34}}(\sigma_{44}{\Omega^{\prime}_{33}}+\sigma_{33}{\Omega^{\prime}_{44}}) r^2 \kappa \Omega_{r}+(\sigma_{44}{{\Omega^{\prime}}^2_{33}}+\sigma_{33}{{\Omega^{\prime}}^2_{34}})r^{2}\Omega_{r}^2\big]},
\nonumber \\
\beta_{1} & = &
-2\D{\sigma_{11}\sigma_{22}r^2\big[(\sigma_{44}W_{34}{\Omega^{\prime}_{34}}+\sigma_{33}W_{44}{\Omega^{\prime}_{44}})(p_{r}^{2}+r^{2}\kappa^{2})+
\big((\sigma_{33}W_{44}+\sigma_{44}W_{33}){\Omega^{\prime}_{34}}+
W_{34}(\sigma_{44}{\Omega^{\prime}_{33}}+\sigma_{33}{\Omega^{\prime}_{44}})\big)
r^{2}\kappa\Omega_{r}} \nonumber \\ 
& &
-\D{(\sigma_{44}W_{33}{\Omega^{\prime}_{33}}+\sigma_{33}W_{34}{\Omega^{\prime}_{34}})r^{2}\Omega_{r}^{2}\big]},
\nonumber \\ 
\beta_{0} & = &
\D{r^{2}\big[\sigma_{11}\big(\sigma_{22}\sigma_{33}W_{44}^{2}+\sigma_{44}(\sigma_{22}W_{34}^{2}+\sigma_{33})
\big)(p_{r}^{2}+r^{2}\kappa^{2})-2\sigma_{11}\sigma_{22}W_{34}(\sigma_{33}W_{44}+\sigma_{44}W_{33})
r^{2}\kappa\Omega_{r}+}
\nonumber \\ 
& &
\D{\sigma_{22}\big(\sigma_{11}\sigma_{33}W_{34}^{2}+\sigma_{44}(\sigma_{11}W_{33}^{2}+\sigma_{33})\big)r^2\Omega_{r}^{2}\big].}
\end{eqnarray}
Finally, multiplying both eigenvalues:
\begin{eqnarray}
\label{eq:lambda_r}
\lambda_{r_{1}}\lambda_{r_{2}} & = & \D{\frac{\det{{\bf \sigma}_{w}^0}}{\lambda_{v_{1}}\lambda_{v_{2}}}}.
\end{eqnarray}

\label{lastpage}

\end{document}